\documentclass[a4paper,12pt]{article}
\usepackage{epsfig}\parskip 5pt plus 1pt
\usepackage{amsmath}
\usepackage{amssymb}
\usepackage{amsfonts}
\usepackage{fleqn}
\usepackage{mathrsfs}
\usepackage{cite}

\def\be{\begin{equation}}
\def\ee{\end{equation}}
\def\bea{\begin{eqnarray}}
\def\eea{\end{eqnarray}}

\begin{document}

\bibliographystyle{OurBibTeX}

\begin{titlepage}

 \vspace*{-15mm}
\begin{flushright}
{SHEP--07--15}\\
\today\\
\end{flushright}
\vspace*{5mm}

\begin{center}
{ \bf \Large Planck Scale Unification in a Supersymmetric Standard Model}
\\[8mm]
R.~Howl\footnote{E-mail: \texttt{rhowl@phys.soton.ac.uk}.} and
S.F.~King\footnote{E-mail: \texttt{sfk@hep.phys.soton.ac.uk}.}
\\
{\small\it
School of Physics and Astronomy, University of Southampton,\\
Southampton, SO17 1BJ, U.K.\\[2mm]
}
\end{center}
\vspace*{0.75cm}

\begin{abstract}
\noindent
\end{abstract}

We show how gauge coupling unification near the Planck scale
$M_p\sim 10^{19}$ GeV can be achieved in the framework of
supersymmetry, facilitating a full unification of all forces with
gravity. Below the conventional GUT scale $M_{GUT}\sim 10^{16}$ GeV
physics is described by a Supersymmetric
Standard Model whose particle content is that of
three complete $\mathbf{27}$ representations of the gauge group
$E_6$. Above the conventional GUT scale the gauge group corresponds
to a left-right symmetric Supersymmetric Pati-Salam model, which may
be regarded as a ``surrogate SUSY GUT'' with all the nice features
of $SO(10)$ but without proton decay or doublet-triplet splitting
problems. At the TeV scale the extra exotic states may be discovered
at the LHC, providing an observable footprint of an underlying $E_6$
gauge group broken at the Planck scale. Assuming an additional low
energy $U(1)_X$ gauge group, identified as a non-trivial combination
of diagonal $E_6$ generators, the $\mu$ problem of the MSSM can be
resolved.

\end{titlepage}
\newpage

\section{Introduction}

Gauge coupling unification and the cancellation of quadratic
divergences are two of the most appealing features of supersymmetric
(SUSY) extensions of the standard model (SM) \cite{Chung:2003fi}. It
is well known that the electroweak and strong gauge couplings
extracted from LEP data and extrapolated to high energies using the
renormalisation group (RG) evolution do not meet within the SM.
However, in the framework of the minimal supersymmetric standard
model (MSSM) \cite{1} the couplings converge to a common value at
some high energy scale. This allows one to embed SUSY extensions of
the SM into Grand Unified Theories (GUTs), leading to SUSY GUTs
based on $SU(5)$ or $SO(10)$.

However, despite their obvious attractions, SUSY GUTs face some
serious challenges from the experimental limits on proton decay on
the one hand, and the theoretical requirement of Higgs
doublet-triplet splitting on the other as recently discussed for
example in \cite{Raby:2004br}. Furthermore the unification of gauge
couplings near a conventional GUT scale $M_{GUT}\sim 10^{16}$ GeV
leaves open the question of a full unification of all the forces
with gravity, although this may be achieved in the framework of
string unification, including high energy threshold effects
\cite{Ross:2004mi}.

It was suggested some time ago that one should consider replacing
the SUSY GUT theory by a Pati-Salam gauge group above $M_{GUT}\sim
10^{16}$ GeV, which plays the role of a ``surrogate SUSY GUT''
\cite{King:1994fv}, since there is no proton decay or
doublet-triplet splitting problem in such a theory. In this scheme
the gauge couplings meet at $M_{GUT}\sim 10^{16}$ GeV, as in the
MSSM, and are then held together up to the Planck scale by a
combination of left-right symmetry and carefully selected matter
content chosen so that the $SU(4)_{PS}$ gauge group has the same
beta function as the $SU(2)_L\times SU(2)_R$ gauge couplings
\cite{King:1994fv}. However such ``theoretical tuning'' of the
$SU(4)_{PS}$ and $SU(2)_L\times SU(2)_R$ beta functions appears to
be somewhat contrived.

Recently a so-called Exceptional Supersymmetric Standard Model
(ESSM) has been proposed \cite{King:2005my,King:2005jy}, in which
the low energy particle content consists of three $\mathbf{27}$
representations of the gauge group $E_6$, plus in addition a pair of
non-Higgs doublets $H',\overline{H}'$ arising from incomplete
$\mathbf{27}',\overline{\mathbf{27}}'$ representations. In the ESSM,
gauge coupling unification works even better than in the MSSM
\cite{King:2007uj}. Although the ESSM solves the usual $\mu$ problem
via a singlet coupling to two Higgs doublets, the presence of the
non-Higgs doublets $H',\overline{H}'$ introduces a new $\mu'$
problem since in this case a singlet coupling generating $\mu'$ is
not readily achieved \cite{King:2005my,King:2005jy}. However the
only purpose of including the non-Higgs states $H',\overline{H}'$ is
to help achieve gauge coupling unification at $M_{GUT}\sim 10^{16}$
GeV. This allows the possibility of removing the non-Higgs states
$H',\overline{H}'$ from the spectrum. Of course the question of
gauge coupling unification must then be addressed, which is the
subject of the present paper.

In this paper we consider a similar model to the ESSM but without the
additional non-Higgs doublets $H',\overline{H}'$. Clearly, without
the additional non-Higgs doublets $H',\overline{H}'$, the gauge
couplings will no longer converge at $M_{GUT}\sim 10^{16}$ GeV, or
any other scale, so at first sight this possibility looks
unpromising. However we shall show that, if the theory is embedded
into a Pati-Salam theory at $M_{GUT}\sim 10^{16}$ GeV, then,
remarkably, this leads to a unification of all forces with
gravity close to the Planck scale. In the region between $M_{GUT}$
and $M_p$ there is not a SUSY GUT but a ``surrogate SUSY GUT'' based
on the Pati-Salam gauge group which resolves the proton decay and
doublet-triplet splitting problems of SUSY GUTs, with Planck scale
unification achieved in a more natural way than in
\cite{King:1994fv}.

Unification in supersymmetric models containing one or three
$\mathbf{27}$ representations of the gauge group $E_6$ has recently
been considered in the literature \cite{Kilian:2006hh}. Assuming an
intermediate Pati-Salam gauge group at the scale $10^{15}$ GeV at
which the Standard Model (SM) couplings satisfy $\alpha_1=\alpha_2$,
it was claimed that the resulting Pati-Salam gauge couplings could
subsequently meet at a higher scale about $10^{18}$ GeV
\cite{Kilian:2006hh}. However the condition $\alpha_1=\alpha_2$
cannot be consistently applied at the Pati-Salam breaking scale.
Instead we find the Pati-Salam breaking scale to be about an order
of magnitude larger than the crossing point $\alpha_1=\alpha_2$,
close to $M_{GUT}\sim 10^{16}$ GeV, with full unification close to
$M_p\sim 10^{19}$. Planck scale unification has also been considered
in non-supersymmetric models in \cite{Lykken:1993br}.
In our analysis we shall naively extrapolate
the two-loop RGEs up to $M_p$, although in reality we expect
new physics effects arising from quantum gravity to set in
about an order of magnitude below this.
For example, although the Planck
scale is usually equated with
the Planck mass energy scale \(M_p\) given by \(M_p =
\sqrt{\hbar c / G} \approx 1.2 \times 10^{19}~ \mathrm{GeV / c^2} \)
where \(G\) is Newton's constant, the scale at which quantum gravity
becomes relevant may be considered to be
$\left(8\pi G \right)^{-1/2}\approx 2.4\times
10^{18}$ GeV, where the factor of $8\pi $ comes from the Einstein
field equation $G^{\mu \nu}=8\pi G T^{\mu \nu}$,
which is sometimes referred to as the reduced Planck
scale. It is around this energy scale
that an effective quantum field theory of gravity is expected to
break down and some new physics takes over since effective quantum
field theories of gravity contain corrections to the predictions of
General Relativity proportional to powers of \(E^2 / M^2_p\) where
\(E\) is the energy scale of interest. A more precise estimate
of the energy scale at which new physics associated with quantum gravity
takes over, based on unitarity violation, may be found in \cite{Han:2004wt},
and we return to this point later.


The layout of the rest of this paper is as follows. In the section 2
we consider the pattern of symmetry breaking assumed in this paper.
In section 3 we consider the two loop RG evolution of gauge
couplings in this model from low energies, through the Pati-Salam
breaking scale at $M_{GUT}\sim 10^{16}$ GeV, assuming various
Pati-Salam breaking Higgs sectors, and show that the Pati-Salam
gauge couplings converge close to the Planck scale $M_p\sim 10^{19}$
GeV. In section 4 we shall construct an explicit supersymmetric
model of the kind we are considering.
Finally we conclude the paper in section 5.

\section{Pattern of Symmetry Breaking}
The two step pattern of gauge group symmetry breaking we analyse in
this paper is:
\begin{equation} \label{eq:steps}
E_6 \overbrace{\longrightarrow}^{M_p} G_{422} \otimes D_{LR}
\overbrace{\longrightarrow}^{M_{GUT}} G_{321}
\end{equation}
where the gauge groups are defined by: \be G_{422} \equiv SU(4)
\otimes SU(2)_L \otimes SU(2)_R, \ \ \ \ G_{321} \equiv SU(3)_c
\otimes SU(2)_L \otimes U(1)_Y \ee and we have assumed that the
first stage of symmetry breaking happens close to the Planck scale
and the second stage happens close to the conventional GUT scale.
The first stage of symmetry breaking is based on the maximal
\(E_6\) subgroup \(SO(10) \otimes U(1)_{\psi}\) and the maximal
\(SO(10)\) subgroup $G_{422} \otimes D_{LR}$ corresponding to a
Pati-Salam symmetry with \(D_{LR}\) being a discrete left-right
symmetry. \footnote{Under \(D_{LR}\) the matter multiplets transform
as \(q_L \rightarrow q^c_L\), and the gauge groups $SU(2)_L$ and
$SU(2)_R$ become interchanged \cite{Chang:1984uy}.}

The pattern of symmetry breaking assumed in this paper is different
from that commonly assumed in the literature based on the maximal
\(SO(10)\) subgroup $SU(5)\otimes U(1)_{\chi}$
\cite{King:2005my,King:2005jy,Hewett:1988xc}. In particular the
Pati-Salam subgroup does not contain the Abelian gauge group factor
$U(1)_{\chi}$. The only Abelian gauge group factor involved in this
pattern of symmetry breaking is $U(1)_{\psi}$, and in the present
analysis we assume that this is broken at $M_p$. However, as
discussed in section 4, there are good phenomenological motivations,
related to the solution to the $\mu$ problem, for preserving a low
energy $U(1)'$ gauge group, and this would require the $U(1)_{\psi}$
gauge group to be preserved. In the present paper we do not consider
the effect of including the $U(1)_{\psi}$ gauge group factor in the
RG analysis, however we have checked that Planck scale unification would still be possible, so the results presented here
would not be much affected by its inclusion.\footnote{With \(U(1)_{\psi}\) included in the RG analysis for the \(\mathbf{27_H}
 + \mathbf{\overline{27}}_{\mathbf{H}}\) graph (right panel of figure 1) it may be necessary to increase the effective MSSM threshold to \(350\) GeV to ensure Planck scale unification for the larger experimental values of the strong coupling constant.}

The first stage of symmetry breaking close to $M_p$ will not be
considered in this paper. We only remark that the Planck scale
theory may or may not be based on a higher dimensional string
theory. Whatever the quantum gravity theory is, it will involve some
high energy threshold effects, which will depend on the details of
the high energy theory, and which we do not consider in our
analysis.

The second stage of symmetry breaking close to $M_{GUT}$ is within
the realm of conventional quantum field theory, and requires some
sort of Higgs sector, in addition to the assumed matter content of
three $\mathbf{27}$ representations of the gauge group $E_6$. In
order to break the Pati-Salam symmetry \(G_{422}\) to \(G_{321}\) at
$M_{GUT}$ the minimal Higgs sector required are the $G_{422}$
representations \(H_R=(4,1,2)\) and
$\overline{H}_R=(\overline{4},1,\overline{2})$ \cite{King:1997ia}. When these
particles obtain VEVs in the right-handed neutrino directions they
break the \(SU(4) \otimes SU(2)_R\) symmetry to \(SU(3)_c \otimes
U(1)_Y\) with the desired hypercharge assignments, as discussed
later.

Although a Higgs sector consisting of \(H_R\) and \(\overline{H}_R\)
is perfectly adequate for breaking Pati-Salam symmetry, it does not
satisfy $D_{LR}$. We must therefore also consider an extended Higgs
sector including their left-right symmetric partners. A minimal
left-right symmetric Higgs sector capable of breaking Pati-Salam
symmetry consists of the \(SO(10) \) Higgs states \(\mathbf{16_H}\)
and \(\overline{\mathbf{16}}_{\mathbf{H}}\). If complete \(E_6\) multiplets are
demanded in the entire theory below $M_p$, then the Pati-Salam
breaking Higgs sector at $M_{GUT}$ may be assumed to be
\(\mathbf{27_H}\) and \(\overline{\mathbf{27}}_{\mathbf{H}}\). Therefore in our
analysis we shall consider two possible Higgs sectors which
contibute to the SUSY beta functions in the region between $M_{GUT}$
and $M_p$, namely either \(\mathbf{16_H}+\overline{\mathbf{16}}_{\mathbf{H}}\)
or \(\mathbf{27_H}+\overline{\mathbf{27}}_{\mathbf{H}}\), where it is
understood that only the Pati-Salam gauge group exists in this
region, and these Higgs representations must be decomposed under the
Pati-Salam gauge group. No such Higgs sectors were included in the
analysis in \cite{Kilian:2006hh}.

When \(H_R\) and $\overline{H}_R$ (contained in either
\(\mathbf{16_H}+\overline{\mathbf{16}}_{\mathbf{H}}\) or
\(\mathbf{27_H}+\overline{\mathbf{27}}_{\mathbf{H}}\)) develop VEVs
in the right-handed neutrino directions they break the \(SU(4)
\otimes SU(2)_R\) symmetry to \(SU(3)_c \otimes U(1)_Y\) with the
desired hypercharge assignments. Six of the \(SU(4)\) and two of the
\(SU(2)_R\) fields are then given masses related to the VEV of the
Higgs bosons and the gauge bosons associated with the \(T^{15}\) and
\(T^3_R\) generators are rotated by the Higgs bosons to create one
heavy gauge boson and the bino gauge boson associated with
\(U(1)_Y\). In breaking \(SU(4) \otimes SU(2)_R\) to \(SU(3)_c
\otimes U(1)_Y\) the SM hypercharge generator is a combination of
the diagonal generator \(T^{15}=\sqrt{\frac{3}{2}}~
diag(\frac{1}{6},\frac{1}{6},\frac{1}{6},-\frac{1}{2})\) of
\(SU(4)\) and the diagonal generator of \(SU(2)_R\),
\(T^3_R=\frac{1}{2}~ diag(1,-1)\). $T^{15}=\sqrt{\frac{3}{2}}(B -
L)/2$ where \(B\) and \(L\) are the baryon and lepton number
assignments of each standard model particle. Comparing these
diagonal generators to the hypercharge values we must have $Y =
T^3_R+(B - L)/2$. Then one finds the following relation between the
hypercharge gauge coupling constant \(g_Y\) and the \(SU(4)\) and
\(SU(2)_R\) gauge coupling constants \(g_4\) and \(g_{2R}\)
respectively \cite{Pati:1974yy}:
\begin{equation} \label{eq:relation}
\frac{1}{\alpha_Y} = \frac{1}{\alpha_{2R}} + \frac{1}{\frac{3}{2}
\alpha_{4}}
\end{equation}
where \(\alpha_Y \equiv \frac{g^2_Y}{4\pi},~\alpha_{2R} \equiv
\frac{g^2_{2R}}{4\pi}\) and \(\alpha_4 \equiv \frac{g^2_4}{4\pi}\).

Because the Pati-Salam symmetry, and hence the standard model, is
assumed to come from an \(E_6\) group, then all the charges and
generators should be correctly normalized.\footnote{We choose to
normalize the \(E_6\) generators \(G^a\) by \(Tr(G^a G^b) = 3
\delta^{a b}\). It then follows that
the Pati-Salam and standard model operators are conventionally
normalized by \(Tr(T^a T^b) = \frac{1}{2} \delta^{ab}\).} In this
case the conventional standard model hypercharge assignments must be
modified by a factor of \(\sqrt{\frac{5}{3}}\).  Therefore
Eq.\ref{eq:relation} should be rewritten in terms of the `GUT'
normalized hypercharge \(g_1 \equiv \sqrt{\frac{5}{3}} g_Y\):
\begin{equation} \label{eq:GUTrelation}
\frac{5}{\alpha_1} = \frac{3}{\alpha_{2R}} + \frac{2}{\alpha_{4}}
\end{equation}
where \(\alpha_1 \equiv \frac{g^2_1}{4\pi}\).
Eq.\ref{eq:GUTrelation} is the boundary condition for the gauge
couplings at the Pati-Salam symmetry breaking scale, in this case
$M_{GUT}$. Due to left-right symmetry, at the Pati-Salam symmetry
breaking scale we have the additional boundary condition
$\alpha_{2L}=\alpha_{2R}$. In \cite{Kilian:2006hh} it was assumed
that at the Pati-Salam symmetry breaking scale
$\alpha_1=\alpha_{2L}=\alpha_{2R}$ which disagrees with
Eq.\ref{eq:GUTrelation}, since $\alpha_4\neq
\alpha_{2L}=\alpha_{2R}$ at this scale, as discussed in the next
section.

\section{Two-loop RG Analysis: Planck Scale Unification}

\begin{figure}
\includegraphics[angle=0, scale=1.012]{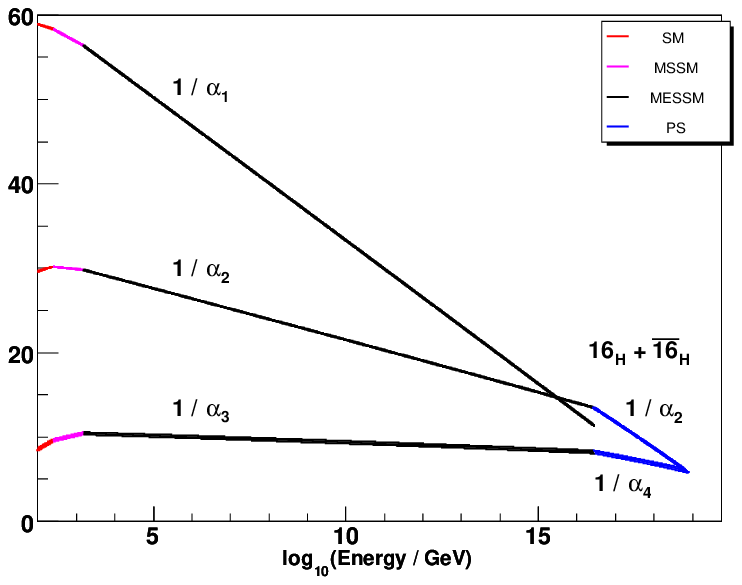} 
  \hspace{0.28 cm}
  \includegraphics[angle=0, scale=1.012]{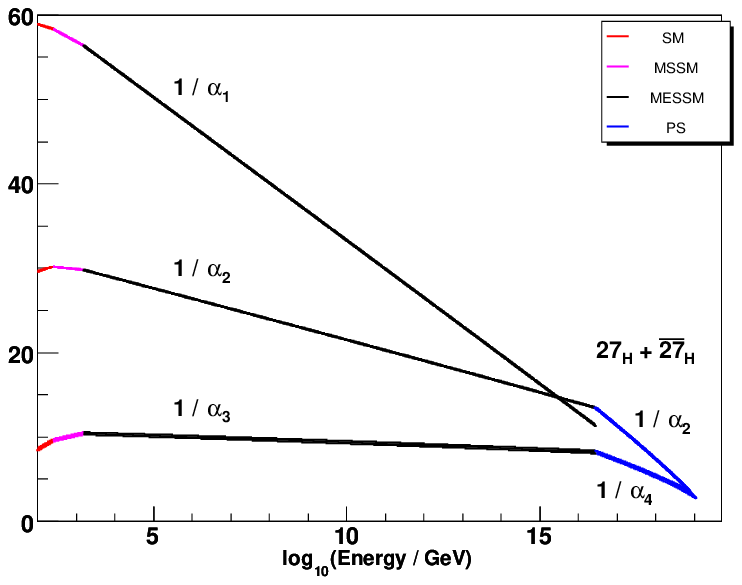} 
\caption{\footnotesize Two-loop Planck Scale Unification in
a supersymmetric model which
contains three generations of SUSY \(\mathbf{27} \)
  particles, above an assumed threshold scale of
  \(1.5\) TeV. Below this scale the MSSM is assumed with a threshold scale
of \(250\) GeV, below which the SM is assumed.
At the scale \(M_{GUT}=10^{16.44(2)}\) GeV
the spectrum is embedded into a left-right symmetric Pati-Salam theory,
with $\alpha_4=\alpha_3$ at \(M_{GUT}=10^{16.44(2)}\) GeV,
and $\alpha_{2L}=\alpha_{2R}$ above this scale.
The left panel contains the additional the SUSY Higgs
contained in \(\mathbf{16_H}\) +
  \(\overline{\mathbf{16}}_{\mathbf{H}}\) while the right
  panel contains the SUSY Higgs in \(\mathbf{27_H}\) +
  \(\overline{\mathbf{27}}_{\mathbf{H}}\)
which enter above \(M_{GUT}=10^{16.44(2)}\) GeV. The Pati-Salam
gauge couplings $\alpha_4$ and $\alpha_{2L}=\alpha_{2R}$
converge at \(10^{18.83(7)}\) GeV and \(10^{18.97(9)}\) GeV for
  the left and right panels respectively, close to the Planck scale,
leading to a unified coupling of \(\alpha_P = 0.166(7)\) or \(\alpha_P
= 0.321(46)\). The numbers in parentheses represent the error
resulting from the experimental error in the coupling constants.
}
\label{b}
\end{figure}

In this section we perform a SUSY two-loop RG analysis of the gauge
couplings, corresponding to the pattern of symmetry breaking
discussed in the previous section. According to our assumptions there are
three complete $\mathbf{27}$ SUSY representations of the gauge group
$E_6$ in the spectrum which survive down to low energies, but,
unlike the original ESSM, there are no additional $H',\overline{H}'$
states so the gauge couplings are not expected to converge at
$M_{GUT}$. Instead, we envisage the pattern of symmetry breaking
shown in Eq.\ref{eq:steps}, where above the Pati-Salam symmetry
breaking scale $M_{GUT}$ we assume, in addition to the three
$\mathbf{27}$ representations, a Pati-Salam symmetry breaking
Higgs sector of either \(\mathbf{16_H}+\overline{\mathbf{16}}_{\mathbf{H}}\) or
\(\mathbf{27_H}+\overline{\mathbf{27}}_{\mathbf{H}}\)
which are assumed to gain masses of order the
Pati-Salam symmetry breaking scale $M_{GUT}$, leaving only the three
$\mathbf{27}$ matter representations below this scale.

For the present RG analysis, we run the couplings up from low
energies to high energies, using as input the SM couplings measured
on the Z-pole at LEP, which are as follows \cite{Yao:2006px}:
\(\alpha_1 (M_Z) = 0.016947(6)\), \(\alpha_2 (M_Z) = 0.033813(27)\)
and \(\alpha_3(M_Z) = 0.1187(20)\). The general two-loop beta
functions used to run the gauge couplings can be found in
\cite{two-loop}.  From $M_Z$ up to an assumed MSSM threshold energy
of 250 GeV we consider only the non-SUSY SM spectrum including a top
quark threshold at 172 GeV. From 250 GeV to 1.5 TeV we include all
the states of the MSSM. From 1.5 TeV up to the Pati-Salam symmetry
breaking scale we include all the remaining states which fill out
three complete SUSY $\mathbf{27}$ representations. The assumed
threshold energies correspond to those in \cite{King:2007uj}, where
a full discussion of MSSM and ESSM threshold effects is given. The
only difference is that here we do not include the
$H',\overline{H}'$ states of the ESSM, so the gauge couplings do not
converge at $M_{GUT}$. Instead $M_{GUT}$ is taken to be the
Pati-Salam symmetry breaking scale, which is determined as follows.

In the previous section we discussed the relation in
Eq.\ref{eq:GUTrelation} between the hypercharge and Pati-Salam
coupling constants at the Pati-Salam symmetry breaking scale. This
can be turned into a boundary condition involving purely \(G_{321}\)
couplings constants at the Pati-Salam breaking scale, since
\(SU(3)_c\) comes from \(SU(4)\) so \(\alpha_3 = \alpha_4\) at this
scale, and, as remarked, \(D_{LR}\) symmetry requires that
\(\alpha_{2R} =\alpha_{2L}\) at the Pati-Salam symmetry breaking
scale. Therefore Eq.\eqref{eq:GUTrelation} can be re-expressed as:
\begin{equation} \label{eq:GUTrelation2}
\frac{5}{\alpha_1} = \frac{3}{\alpha_{2L}} + \frac{2}{\alpha_{3}}.
\end{equation}
Having specified the low energy matter content, and thresholds,
Eq.\ref{eq:GUTrelation2} allows a unique determination of the
Pati-Salam breaking scale, by simply running up the gauge couplings
until the condition is satisfied. In practice, ${\alpha_{3}}$ runs
quite slowly (its one loop beta-function is zero), while the inverse
hypercharge coupling decreases most rapidly and the condition is
satisfied for a Pati-Salam symmetry breaking scale about an order of
magnitude higher energy scale than the crossing point of
${\alpha_{1}}$ and ${\alpha_{2}}$ assumed in \cite{Kilian:2006hh}.
Assuming the above matter content and threshold corrections, the
Pati-Salam symmetry is found to be broken at \(M_{GUT}=10^{16.44(4)}\)
GeV as illustrated in Figure 1. This is close to the conventional
GUT energy scale, and justifies our use of the notation $M_{GUT}$ to
denote the Pati-Salam breaking scale.

Above the scale $M_{GUT}$ we run up the two Pati-Salam gauge
couplings, namely $\alpha_4$ and $\alpha_{2L}=\alpha_{2R}$,
including, in addition to the three SUSY $\mathbf{27}$ matter
representations, also a Pati-Salam SUSY Higgs breaking sector
consisting of either \(\mathbf{16_H}+\overline{\mathbf{16}}_{\mathbf{H}}\) or
\(\mathbf{27_H}+\overline{\mathbf{27}}_{\mathbf{H}}\). The Pati-Salam couplings
are found to converge at at either \(10^{18.83(7)}\) GeV or
\(10^{18.97(9)}\) GeV, respectively, as shown in Figure 1.\footnote{If we were to drop the \(D_{LR}\)
symmetry then, with a minimal Pati-Salam Higgs content consisting of
just \(H_R\) and $\overline{H}_R$, the equation \(\frac{5}{\alpha_1}
= \frac{3}{\alpha_{2R}} + \frac{2}{\alpha_3}\) at the Pati-Salam
scale would predict that the Pati-Salam symmetry is broken at
\(10^{14.4(1)}\) GeV and that unification would occur at
\(10^{19.72(15)}\) GeV.}
These values are
close to the Planck scale $M_p=1.2\times 10^{19}$ GeV, and suggests
a Planck scale unification of all forces with gravity.

The value of the gauge coupling constant at the unification scales
\(10^{18.83(7)}\) GeV or \(10^{18.97(9)}\) GeV is \(\alpha_P =
0.166(7)\) or \(\alpha_P = 0.321(46)\) for the \(\mathbf{16}_H +
\overline{\mathbf{16}}_\mathbf{H}\) or \(\mathbf{27}_\mathbf{H} +
\overline{\mathbf{27}}_\mathbf{H}\) particle spectra, respectively.
These values of the unified gauge coupling at the Planck scale are
much larger than the conventional values of $\alpha_{GUT}$, and
indeed are larger even than $\alpha_3(M_Z)$, however they are still
in the perturbative regime. Of course there are expected to be large
threshold corrections coming from Planck scale physics which are not
included in our analysis. Indeed, we would expect that QFT breaks
down as we approach the Planck scale, so that our RG analysis ceases
to be valid as we approach the Planck scale, as remarked in the
Introduction. The precise energy scale \(E_{new}\) at which quantum field
theories of gravity are expected to break down and new physics takes
over is discussed in \cite{Han:2004wt} based on estimates of the
scale of violation of (tree-level) unitarity.
An upper bound for this new physics energy scale
is given by \(E^2_{new} = 20 [G (\frac{2}{3} N_s + N_f + 4N_V)]^{-1}
\) where \(N_s\), \(N_f\) and \(N_V\) are the number of scalars,
fermions and vectors respectively that gravity couples to. Assuming three
low-energy \(\mathbf{27}\) multiplets, \(E_{new}\) would be equal to
\(10^{18.6}\) GeV which sets an upper bound for the scale at which
our quantum field theory analysis (and with any corrections from
effective quantum gravity theories included) can no longer be
trusted. We have shown that the gauge coupling constants are
predicted to be very close to one another at this scale and that, if
extrapolated, unify just below \(M_p\). We have naively
extrapolated the RGEs up to \(M_p\), even though new physics
associated with quantum gravity must enter an order of magnitude
below this. The fact that the two PS couplings are
very close to each other at \(E_{new}\), and are on a
convergent trajectory must be regarded, at best, as a suggestive hint of a
unification of the gauge fields with gravity in this approach.

\section{Constructing a Realistic Model}

\begin{table}
\begin{center}
 \begin{tabular}{ | c | c |c | c| }
 \hline
  \(Z_2\) & \(R\)-charges &
  Incomplete $E_6$ multiplets
    \\ \hline \(+\) &\(\frac{2}{3}\) & \(\phi_0 = S_3,~h_{3}\)
 \\
\hline
\( - \)& \(\frac{2}{3}\) & \(\phi_i = F_i,~F^c_i,~\mathcal{D}_{i},~h_{\alpha},~S_{\alpha}\)
\\ \hline
\( + \)& \(\frac{1}{3}\) & \(\mathbf{16}_\mathbf{H}\) or
\(\mathbf{27}_\mathbf{H}\)
\\ \hline
\( + \)& \(\frac{1}{3}\) & \(\overline{\mathbf{16}}_\mathbf{H}\) or
\(\overline{\mathbf{27}}_\mathbf{H}\)
\\ \hline
\( - \)& \(0\) & \(\Sigma \)
\\ \hline
\( + \)& \(\frac{4}{3}\) & \(M \)
\\ \hline
\end{tabular}
\end{center}
\caption{\footnotesize The \(Z_2\) and R-charge assignments of the
Pati-Salam respecting incomplete \(E_6\) multiplets of the MESSM.
The four \(\mathbf{27}\) multiplets are divided into \(\phi_0\) and
\(\phi_i\) where \(i = 1 \ldots 3\). \(\phi_0\) contains the MSSM
Higgs doublets \(h_3\) and a Pati-Salam symmetry singlet \(S_3\)
which gives mass to the \(\mathcal{D}_i\) and \(h_{\alpha}\)
particles in \(\phi_i\) where \(\alpha = 1,2\). \(\phi_i\) contains
the quarks and leptons \(F_i\) and \(F^c_i\) as well as the exotic
quarks \(\mathcal{D}_i\) and non-Higgs \(h_{\alpha}\).  The
\(\mathbf{16}_\mathbf{H}\) and \(\overline{\mathbf{16}}_\mathbf{H}\)
or \(\mathbf{27}_\mathbf{H}\) and
\(\overline{\mathbf{27}}_\mathbf{H}\) multiplets of \(SO(10)\) and
\(E_6\) respectively break the Pati-Salam symmetry and are given a
GUT scale mass from the \(E_6\) singlet M. \(\Sigma\) is another
\(E_6\) singlet that gets a VEV within the energy range
\(10^{7-11}\)GeV to sufficiently suppress proton decay and to
allow the exotic particles to decay with a rate that avoids
cosmological problems.} \label{b}
\end{table}

We have shown that a low energy matter and Higgs content corresponding
to three \(\mathbf{27}\) multiplets of the \(E_6\),
embedded in a left-right Pati-Salam symmetry at the GUT scale,
can lead to Planck scale unification. However, assuming
superpotential interactions corresponding to
the \(E_6\) respecting operator \(\mathbf{27}_i \mathbf{27}_j
\mathbf{27}_k\) where \(i,~j\) and \(k\) are family indices, leads
to trouble with phenomenology due to proton decay mediated
by TeV scale colour triplets, and flavour changing neutral currents
(FCNCs) due to multiple Higgs-like doublets. The solution proposed in
\cite{King:2005jy,King:2007uj} is to consider incomplete
\(E_6\) supermultiplets at low energies, which allows extra
symmetries to appear at low energies that forbid proton
decay and suppress FCNCs, while allowing the exotic
colour triplets to decay with a lifetime less than about
one second, to avoid conflict with nucleosynthesis.
The same symmetries cannot be used here
since they would not respect the Pati-Salam symmetry, so we must therefore
seek alternative symmetries that are consistent with the Pati-Salam symmetry.

We now propose a realistic supersymmetric model which contains the
matter content of three $\mathbf{27}$'s of $E_6$ at low energy, 
arising from four incomplete $\mathbf{27}$'s of $E_6$ at high energy, 
and is embeddable in Pati-Salam (PS), without leading to conflict with
phenomenology or cosmology. We shall call such a model the minimal
exceptional supersymmetric standard model (MESSM) to distinguish it
from the ESSM. In the MESSM we shall impose a $Z_2\otimes U(1)_R$
symmetry, under which incomplete multiplets of four \(E_6\)
\(\mathbf{27}\) states \(\phi_0\) and \(\phi_i\), where \(i = 1
\ldots 3\), transform as shown in Table 1. As in the ESSM, we assume
that two Higgs doublets which transform under the PS symmetry as
$h_3=(1,2,2)$ and Higgs singlet $S_3=(1,1,1)$ arise from the
$\phi_0$ representation, which replace the Higgs in the $\phi_3$,
leaving the corresponding states $h_{\alpha}=(1,2,2)$ and
$S_{\alpha}=(1,1,1)$ in the $\phi_{1,2}$ representations which do
not develop VEVs or couple to quarks or leptons, and are
``non-Higgs''.  The remaining matter content of the $\phi_i$,
consist of usual quarks and leptons which transform under PS as
$F_i=(4,1,2)$ and $F_i^c=(\bar{4},1,2)$, the Pati-Salam singlets
\(S_{\alpha}=(1,1,1)\) and coloured vector-like quarks
$\mathcal{D}_i=D_i + \overline{D}_i=(6,1,1)$.  Interactions between
the non-Higgs and matter particles and exotic quarks and matter
particles from the \(\mathbf{\phi}_i \mathbf{\phi}_j
\mathbf{\phi}_k\) operator would introduce flavour changing neutral
currents and rapid proton decay respectively.  We suppress these
phenomenologically problematic terms by the \(Z_2\) symmetry.
However, forbidding all the interactions between the exotic quarks
and matter particles could cause serious cosmological problems since
this would ensure that the lightest exotic quark is a stable
particle whereas cosmology requires the lightest exotic to decay
with a lifetime less than about \(0.1\)s to avoid any problems with
nucleosynthesis \cite{Kawasaki:2004qu}. Therefore the \(Z_2\) symmetry must be
broken by the VEV of a new singlet $\Sigma$ which is sufficiently
large to enable the exotic quarks to decay rapidly, but sufficiently
small to be consistent with proton decay. This turns out to be
possible as we discuss shortly. As well as the three incomplete
\(\mathbf{27}\) multiplets of \(E_6\), there are additional
\(\mathbf{16}_\mathbf{H} + \overline{\mathbf{16}}_\mathbf{H}\) or
\(\mathbf{27}_\mathbf{H} + \overline{\mathbf{27}}_\mathbf{H}\) Higgs
multiplets that break the Pati-Salam symmetry and are assumed to
reside at the GUT scale.  To give a GUT scale mass to these Higgs
multiplets we introduce another \(E_6\) singlet \(M\), which gets a
VEV at the GUT scale.

Using Table 1 the only superpotential terms
that are allowed are the following:
\(\phi_0 \phi_0 \phi_0\) which
generates the term $S_3h_3h_3$ which gives an
effective $\mu$-term when \(S_3\) gets a VEV;
 \(\phi_0 \phi_i \phi_j\) which generates the rest of the MSSM superpotential
terms and gives mass to the ``non-MSSM'' particles
\(\mathcal{D}_i\), \(h_{\alpha}\) and \(S_{\alpha}\);
\(\frac{1}{M_p} \Sigma \phi_i \phi_j \phi_k\) which allows the
exotic quarks to decay once \(\Sigma\) gets a VEV at an energy scale
as discussed below; \(\frac{1}{M_p} \Sigma \phi_i \phi_0 \phi_0\)
which gives a mass mixing between Higgs and non-Higgs;
\(\frac{1}{M_p} \phi_i \phi_j \overline{\mathbf{16}}_\mathbf{H}
\overline{\mathbf{16}}_\mathbf{H}\) (or \(\frac{1}{M_p} \phi_i
\phi_j \overline{\mathbf{27}}_\mathbf{H}
\overline{\mathbf{27}}_\mathbf{H}\)) which generates right-handed
neutrino masses;\footnote{The superpotential term \(\frac{1}{M_p}
\phi_i \phi_j \overline{\mathbf{16}}_\mathbf{H}
\overline{\mathbf{16}}_\mathbf{H}\) is meant to represent all the
Pati-Salam operators that are found once the \(SO(10)\) multiplet
\(\overline{\mathbf{16}}_\mathbf{H}\) is decomposed into its
Pati-Salam representations.  Similar meanings apply to the
superpotential terms \(\frac{1}{M_p} \phi_i \phi_j
\overline{\mathbf{27}}_\mathbf{H}
\overline{\mathbf{27}}_\mathbf{H}\), \(\frac{1}{M_p} \phi_0 \phi_0
\overline{\mathbf{16}}_\mathbf{H}
\overline{\mathbf{16}}_\mathbf{H}\), \(\frac{1}{M_p} \phi_0 \phi_0
\overline{\mathbf{27}}_\mathbf{H}
\overline{\mathbf{27}}_\mathbf{H}\),
\(M\mathbf{16}_\mathbf{H}\overline{\mathbf{16}}_\mathbf{H}\) and
\(M\mathbf{27}_\mathbf{H}\overline{\mathbf{27}}_\mathbf{H}\).}
\(\frac{1}{M_p} \phi_0\phi_0 \overline{\mathbf{16}}_\mathbf{H}
\overline{\mathbf{16}}_\mathbf{H}\) (or \(\frac{1}{M_p} \phi_0
\phi_0 \overline{\mathbf{27}}_\mathbf{H}
\overline{\mathbf{27}}_\mathbf{H}\)) which are harmless; and
\(M\mathbf{16}_\mathbf{H}\overline{\mathbf{16}}_\mathbf{H}\) (or
\(M\mathbf{27}_\mathbf{H}\overline{\mathbf{27}}_\mathbf{H}\)) which
gives a GUT scale mass to the Higgs multiplets \(\mathbf{16_H} +
\overline{\mathbf{16}}_{\mathbf{H}}\) or \(\mathbf{27_H} +
\overline{\mathbf{27}}_{\mathbf{H}}\).

The \(D_i + \overline{D}_i\) components of the superpotential term
\(\phi_i \phi_j \phi_k\) cause proton decay through the decay
channels \(p \rightarrow K^{+} \overline{\nu}\) via \(d = 5\)
operators (via the \(\phi_0 \phi_j \phi_k\) term which is
responsible for the triplet mass $m_D$) and \(p \rightarrow \pi^0
e^{+}\) via \(d = 6\) operators with matrix elements proportional to
\(1 / m_D\) and \(1 / m^2_D\) respectively. If the exotic quarks get
a mass of order \(m_D=1.5\) TeV from \(\phi_0 \phi_j \phi_k\) then
we estimate that the term \(\phi_i \phi_j \phi_k\) must be
multiplied by an effective Yukawa coupling smaller than about
\(10^{-8}\) for the proton's lifetime to be above \(1.6 \times
10^{33}\) years and \(5.0 \times 10^{33}\) years which are the
present experimental limits for the \(p \rightarrow K^{+}
\overline{\nu}\) and \(p \rightarrow \pi^0 e^{+}\) decay modes
respectively \cite{Raby}.  In the MESSM the \(\phi_i \phi_j
\phi_k\) terms are forbidden by a \(Z_2\) symmetry but are
effectively generated from the non-renormalizable terms
\(\frac{1}{M_p} \Sigma \phi_i \phi_j \phi_k\) when \(\Sigma\) gets a
VEV which must therefore be less than \(10^{11}\) GeV to avoid
experimentally observable proton decay.  The effective operators
\(\phi_i \phi_j \phi_k\) from \(\frac{1}{M_p} \Sigma \phi_i \phi_j
\phi_k\) are also the source of exotic quark decay in this model
and, for the exotic quarks to have a lifetime less than \(0.1\)s, we
estimate that the \(\phi_i \phi_j \phi_k\) operators must be
multiplied by an effective Yukawa coupling no less than
\(10^{-12}\), in which case \(\Sigma\) must get a VEV greater than
\(10^{7}\) GeV. Therefore, for the model to be phenomenologically
acceptable, we require that the \(E_6\) singlet \(\Sigma\) should
get a VEV between \(10^{7-11}\) GeV.


As well as the \(Z_2\) and \(U(1)_R\) symmetries, a \(U(1)_{\psi}\)
symmetry from \(E_6\) may also be assumed.
In our RG analysis we
assumed for simplicity that \(U(1)_{\psi}\) is broken at
$M_p$. However
\(U(1)_{\psi}\) may be unbroken at $M_p$, and is broken instead at
$M_{GUT}$ by the Pati-Salam symmetry breaking Higgs \(H_R\) and
$\overline{H}_R$. These Higgs also break the diagonal generator
$T^3_R$ of $SU(2)_R$ and the $B-L$ generator of $SU(4)$ down to
hypercharge $Y$, with $Y = T^3_R+(B - L)/2$ taking a zero value for
the right-handed neutrino and anti-neutrino components. However this
is not the only Abelian generator that is preserved by this Higgs
sector. Under \(E_6\rightarrow SO(10) \otimes U(1)_{\psi}\),
$\mathbf{27}\rightarrow \mathbf{16_{1/2}+10_{-1}+1_{2}}$, and the
\(U(1)_{\psi}\) generator may be written as
$T_{\psi}=diag(1/2,-1,2)$.\footnote{The respective \(E_6\) normalized
generator is \(G^{78} = \frac{1}{\sqrt{6}} diag(\frac{1}{2}, -1,2)\).}
The right-handed neutrino component of the Higgs sector which develops
the VEV will therefore also preserve the generators $T_{\psi}-(B -
L)/2$ and $T_{\psi}+T^3_R$, in addition to $Y = T^3_R+(B - L)/2$. It
is straightforward to show that precisely one additional Abelian
generator orthogonal to $U(1)_Y$ is preserved, namely: \be \label{X}
X=(T_{\psi}+T^3_R)-c_{12}^2Y \ee where $c_{12}=\cos \theta_{12}$ and
the mixing angle is given by \be \tan
\theta_{12}=\frac{g_{2R}}{g_{B-L}}, \ \ \ \
g_{B-L}=\sqrt{\frac{3}{2}}\ g_4, \ee where the Pati-Salam coupling
constants ${g_{2R}}$ and ${g_{4}}$ are evaluated at $M_{GUT}$.

The $U(1)_X$ associated with the preserved generator in Eq.\ref{X}
is an anomaly-free gauge group which plays the same role in solving
the $\mu$ problem as the $U(1)_N$ of the ESSM, since it allows the
coupling $Sh_uh_d$ which generates an effective $\mu$ term, while
forbidding $S_3$ and the $\mu h_uh_d$. $U(1)_X$ is broken by the $S$
singlet VEV near the TeV scale, yielding a physical $Z'$ which may
be observed at the LHC. We emphasize that this $Z'$ is distinct from
those usually considered in the literature based on linear
combinations of the $E_6$ subgroups \(U(1)_{\psi}\) and
\(U(1)_{\chi}\) since, in the MESSM, \(U(1)_{\chi}\) is necessarily
broken at $M_p$. In particular the $Z'$ of the MESSM based on
$U(1)_X$ and that of the ESSM based on $U(1)_N$ will have different
physical properties.

\section{Summary}
In this paper we have proposed and discussed a supersymmetric
standard model, valid below the conventional GUT scale
$M_{GUT}\sim 10^{16}$ GeV. The particle content
consists of three complete SUSY
$\mathbf{27}$ representations of $E_6$. However the $E_6$ gauge
group is broken at $M_p$ and the low energy gauge group is just that
of the SM, supplemented by an additional $U(1)_X$ gauge group. The
Higgs doublets which break electroweak symmetry arise
from a $\mathbf{27}$ representation of $E_6$ as in the ESSM. The
model is an example of a low energy Supersymmetric Standard Model (SSM)
whose spectrum contains only complete GUT representations. In this
respect it is quite unlike any of the conventional SSM's in
the literature such as the MSSM, NMSSM or ESSM which all contain low
energy Higgs (or non-Higgs) states which do not form complete GUT
representations.

We have shown that while gauge coupling at $M_{GUT}\sim 10^{16}$ GeV
is lost, the new model suggests a gauge coupling unification
near the Planck scale $M_p\sim 10^{19}$ GeV. Although we have
naively extrapolated the RGEs up to $M_p$, we have
discussed the fact that in reality there will be new physics effects
associated with quantum gravity that will enter about an
order of magnitude below this scale, so the idea of Planck
scale unification must be considered with caution. Therefore
our results can only be regarded as suggestive of Planck scale unification.
All we can say is that a supersymmetric model
with the low energy matter content of three $\mathbf{27}$'s of
$E_6$, embedded in a Pati-Salam gauge group above $M_{GUT}$,
gives rise to a high energy theory with two gauge couplings which are
very close to each other and converging at the scale at which
quantum gravity effects are expected to set in.

As remarked, above $M_{GUT}\sim
10^{16}$ GeV the model is embedded into a a left-right symmetric
Supersymmetric Pati-Salam model, which may be regarded as a
``surrogate SUSY GUT'' with all the nice features of $SO(10)$ but
without proton decay or doublet-triplet splitting problems. The
Pati-Salam gauge group is broken at $M_{GUT}\sim 10^{16}$ GeV by a
Higgs sector contained in either
\(\mathbf{16_H}+\overline{\mathbf{16}}_\mathbf{H}\) or
\(\mathbf{27_H}+\overline{\mathbf{27}}_\mathbf{H}\), leaving only
the desired spectrum below this scale. The three right-handed
neutrinos and sneutrinos gain masses set by the scale
$M_{GUT}^2/M_p$. At the TeV scale the extra exotic states of the
model, which fill out three complete SUSY $\mathbf{27}$
representations (minus the three right-handed neutrinos and
sneutrinos) may be discovered at the LHC, providing an observable
footprint of an underlying $E_6$ gauge group broken at the Planck
scale. We have shown that it is possible to construct a realistic
supersymmetric model of this kind, which is consistent with
phenomenology and cosmology, by using incomplete $E_6$ multiplets
and assuming a symmetry $Z_2\otimes U(1)_R$, and we called the
resulting model the MESSM to distinguish it from the ESSM.

In our unification analysis performed in this paper we assumed for
simplicity that the $U(1)_{\psi}$ gauge group is broken at $M_p$.
However, in order to solve the $\mu$ problem, we have shown that it
is necessary that the $U(1)_{\psi}$ gauge group survives down to
$M_{GUT}\sim 10^{16}$ GeV, so that below this scale a low energy
$U(1)_X$ gauge group emerges as a linear combination of
$U(1)_{\psi}$ and diagonal Pati-Salam generators. Although we have
not considered the effect of the $U(1)_{\psi}$ and $U(1)_X$ gauge
groups in the RG analysis, we have checked that Planck scale unification is
still possible if they are included. The
phenomenology of a low energy $Z'$, corresponding to a $U(1)_X$
which is not a simple linear combination of the $E_6$ gauge group
\(U(1)_{\psi}\) and \(U(1)_{\chi}\), has not so far been considered
in the literature and requires a dedicated study \cite{Howl}.

\subsection*{Acknowledgments}
We would like to thank Peter Zerwas for encouragement.
RJH acknowledges a PPARC Studentship.
SFK acknowledges partial support from the following grants:
PPARC Rolling Grant PPA/G/S/2003/00096;
EU Network MRTN-CT-2004-503369;
EU ILIAS RII3-CT-2004-506222;
NATO grant PST.CLG.980066.

\end{document}